\documentclass[a4paper]{IEEEtran} 
\IEEEoverridecommandlockouts
\usepackage{cite}
\usepackage{amsmath,amssymb,amsfonts}
\usepackage{algorithmic}
\usepackage{graphicx}
\usepackage{textcomp}
\def\BibTeX{{\rm B\kern-.05em{\sc i\kern-.025em b}\kern-.08em
    T\kern-.1667em\lower.7ex\hbox{E}\kern-.125emX}}
\begin{document}


\title{BlendSM-DDM: BLockchain-ENabled Secure Microservices for Decentralized Data Marketplaces}

\author{
\IEEEauthorblockN{Ronghua Xu${^a}$, Gowri Sankar Ramachandran$^{b}$, Yu Chen${^a}$, Bhaskar Krishnamachari$^{b}$}

\IEEEauthorblockA{${^a}$Dept. of Electrical \& Computer Engineering, Binghamton University, SUNY,  Binghamton, NY 13902, USA \\ $^{b}$Dept. of Electrical Engineering, University of Southern California, Los Angeles, CA 91801, USA\\
\{rxu22, ychen\}@binghamton.edu, \{gsramach, bkrishna\}@usc.edu}
}

\maketitle

\begin{abstract}
To promote the benefits of the Internet of Things (IoT) in smart communities and smart cities, a real-time data marketplace middleware platform, called the Intelligent IoT Integrator (I3), has been recently proposed. While facilitating the easy exchanges of real-time IoT data streams between device owners and third-party applications through the marketplace, I3 is presently a monolithic, centralized platform for a single community. Although the service oriented architecture (SOA) has been widely adopted in the IoT and cyber-physical systems (CPS), it is difficult for a monolithic architecture to provide scalable, inter-operable and extensible services for large numbers of distributed IoT devices and different application vendors. Traditional security solutions rely on a centralized authority, which can be a performance bottleneck or susceptible to a single point of failure. Inspired by containerized microservices and blockchain technology, this paper proposed a BLockchain-ENabled Secure Microservices for Decentralized Data Marketplaces (BlendSM-DDM). Within a permissioned blockchain network, a microservices based security mechanism is introduced to secure data exchange and payment among participants in the marketplace. BlendSM-DDM is able to offer a decentralized, scalable and auditable data exchanges for the data marketplace.

\end{abstract}


\section{Introduction}
\label{sec:intro}  

The proliferation of the Internet of Things (IoT) technology allows the concept of Smart Cities become feasible, and these IoT based applications have greatly improved the lives of residents through enhancing health, safety, and convenience. However, the first generation of IoT deployment (IoT 1.0) for smart city has come across several challenges that prevent wider adoption of the IoT in smart communities and smart cities \cite{ramachandran2019towards}. To enhance the adoption of the IoT in smart communities and smart cities, a marketplace architecture and middleware platform, called the Intelligent IoT Integrator (I3), is proposed to address key problems in IoT marketplace \cite{krishnamachari2018i3}. To build smart communities, I3 is aimed to build a ''data rivers'' that allow data streams from different entities to be merged together and analyzed, processed, and acted upon as needed to support a diverse set of applications \cite{krishnamachari2018i3}. In a commercial aspect, I3 tries to build a marketplace that allows a large volume of devices owners to sell data streams, while different applications can buy one or more data streams used for their data analytic services. 

While the I3 system facilitates the easy exchange of real-time IoT data streams between devices owners and third-party applications through marketplace, it also brings new concerns on architecture and security issues. The I3 system is deployed in a distributed network environment that includes a large number of IoT devices with high heterogeneity and dynamics and different application vendors using non-standard development technology. It requires a scalable, flexible and lightweight system architecture that supports fast development and easy deployment among participants. In addition, a fundamental principle embodied in the vision of I3 is that device and data owners should have the right to decide when they will share their data (or access to their devices, in case of actuators), with whom, and at what price \cite{krishnamachari2018i3}. Thus, flows of data are mediated through end-to-end agreements between devices owners and application developers, which means that enforcing a centralized management policy on I3 becomes difficult and inefficient. Furthermore, a scalable, flexible and fine-grained data sharing and access control mechanism is needed to protect data exchange and payment among entities.


An earlier work has explored how a decentralized data marketplace (DDM) could be created using blockchain and other distributed ledger technologies \cite{ramachandran2018towards}. The possible benefits of such a decentralized architecture are highlighted and the preliminary work shows how key elements of such a decentralized marketplace could be implemented using smart contracts. Other marketplaces in this space include Ocean Protocol \cite{oceanpocotol} and IOTA Data Marketplace \cite{iota}. One of the challenges in implementing an efficient DDM lies in the management of system complexity. With many possible moving parts and components, the decentralized marketplace system could get fragmented or hard to scale. On the other hand, keeping too rigid a framework could also hurt scalability.

In this position paper, a BLockchain-ENabled Secure Microservices for DDM (BlendSM-DDM) is proposed to secure data exchange among different devices owner and application developers. Leveraging the fine-granularity and loose-coupling features of the microservices architecture \cite{butzin2016microservices, datta2018next}, the BlendSM-DDM decouples the DDM applications and security functionality into multiple containerized microservices that are computationally affordable to each individual devices. The security mechanism is implemented as separated microservices that are built on the smart contract. The distributed microservices could cooperate with each other as a service pool to perform complicated data analytical missions and security enforcement without hurting the flexibility and scalability.

The remainder of this paper is organized as follows: Section \ref{sec:relatedwork} provides the background knowledge and reviews the state of the arts research in blockchain and microservice for IoT-based system. Section \ref{sec:architecture} illustrates the details of core security features introduced by the proposed BlendSM-DDM. Section \ref{sec:conclusion} concludes this paper with our ongoing work.

\section{Related Works}
\label{sec:relatedwork}  

\subsection{Microservices in IoT}
A service-oriented architecture (SOA) is widely adopted in the development of application software in an IoT environment \cite{butzin2016microservices}. The traditional SOA utilizes a monolithic architecture that constitutes different software features in a single interconnected and interdependent application and database. Owing to the tightly coupled dependence among functions and components, such a monolithic framework is difficult to adapt to new requirements in an IoT-enabled system, such as scalability, service extensibility, data privacy, and cross-platform interoperability \cite{datta2018next}. The \textit{microservices architecture} allows functional units of an application to work independently with a loose coupling though encapsulating a minimal functional software module as a microservice, which can be individually developed and deployed. finally, multiple decentralized individual microservices cooperatively perform the functions of complex systems.
The flexibility of microservices enables continuous, efficient, and independent deployment of application function units. As two most significant features of the microservices architecture, \textit{fine granularity} means each of the microservices can be developed in different frameworks and with minimal development resources, while \textit{loose coupling} implies that functions of microservices and its components are independent of each other's deployment and development \cite{yu2018survey}.

Thanks to the fine-granularity and loose-coupling properties, the microservices architecture has been investigated in many smart solutions to enhance the scalability and security of IoT-based applications. The IoT systems are advancing from ``things''-oriented ecosystem to a widely and finely distributed microservices-oriented-ecosystem \cite{datta2018next}. An Intelligent Transportation Systems (ITS) that incorporates and combines the IoT approaches using the serverless microservices architecture has been designed and implemented to help the transportation planning for the Bus Rapid Transit (BRT) systems \cite{herrera2018smart}. To enable a more scalable and decentralized solution for advanced video stream analysis for large volumes of distributed edge devices, a system design of a robust smart surveillance systems was proposed based on microservices architecture and blockchain technology \cite{nagothu2018microservice, nikouei2019decentralized, xu2019blendmas}. It aims at offering a scalable, decentralized and fine-grained access control solution for smart public safety. 

\subsection{Blockchain and Smart Contract}
As a fundamental technology of Bitcoin \cite{nakamoto2008bitcoin}, \textit{blockchain} initially was used to promote a new cryptocurrency that performs commercial transactions among independent entities without relying on a centralized authority, like banks or government agencies. Essentially, the blockchain is a public ledger based on consensus rules to provide a verifiable, append-only chained data structure of transactions. Thanks to the decentralized architecture that does not rely on a centralized authority, blockchain allows the data to be stored and updated distributively. The transactions are approved by miners and recorded in the time-stamped blocks, where each block is identified by a cryptographic hash and chained to preceding blocks in a chronological order. In a blockchain network, a \textit{consensus mechanism} is enforced on a large amount of distributed nodes called miners to maintain the sanctity of the data recorded on the blocks. Thanks to the “trustless” proof mechanism running on miners across the network, users can trust the system of the public ledger stored worldwide on many different decentralized nodes maintained by ''miner-accountants'', as opposed to having to establish and maintain trust with a transaction counter-party or a third-party intermediary \cite{swan2015blockchain}. Thus, blockchain is an ideal decentralized architecture to ensure distributed transactions among all participants in a trustless environment, like edge-based IoT networks.

Emerging from the intelligent property, a \textit{smart contract} allows users to achieve agreements among parties through a blockchain network. By using cryptographic and security mechanisms, a smart contract combines protocols with user interfaces to formalize and secure relationships over computer networks \cite{szabo1997formalizing}. A smart contract includes a collection of pre-defined instructions and data that have been saved at a specific address of blockchain as a Merkle hash tree, which is a constructed bottom-to-up binary tree data structure. Through exposing public functions or application binary interfaces (ABIs), a smart contract interacts with users to offer the predefined business logic or contract agreement.

Other distributed data sharing platforms in this space includes Trinity \cite{ramachandran2019trinity}, which is a distributed pub-sub data broker with support for blockchain-based immutable data storage, Pub-Pay-Sub Protocol \cite{ramachandran2019publish}, which is a publish-subscribe broker with built-in support for micropayments, and SDPP \cite{radhakrishnan2018streaming}, which is a peer-to-peer streaming data protocol with micropayments support.

The blockchain and smart contract enabled security mechanism for applications has been a hot topic and some efforts have been reported recently, for example, smart surveillance system \cite{nagothu2018microservice, nikouei2018realtime, xu2019blendmas}, social credit system \cite{xu2018constructing}, space situation awareness \cite{xu2018exploration}, biomedical imaging data processing \cite{xu2019decentralized}, identification authentication \cite{hammi2018bubbles} and access control \cite{xu2018blendcac, xu2018smartcac}. Blockchain and smart contract together are promising to provide a solution to enable a secured data sharing and access authorization in decentralized IoT systems.

\section{System Architecture of BlendSM-DDM}
\label{sec:architecture}  
In the vision of I3, the system is oriented for device owners and data users, and the data exchanged payment are mainly between sellers and buyers. To improve the economically self-sustaining business ecosystem, blockchain is a very promising solution that will enrich the I3 system by providing a trusted sharing service, where data is reliable, immutable and auditable. The virtualization technology, like virtual machines (VMs) or containers, is platform independent and could provide resource abstraction and isolation features, they are ideal for system architecture design to address the heterogeneity challenge in IoT system. Compared to VMs, containers are more lightweight and flexible with OS-level isolation, so that is an ideal selection for edge computing. 

\begin{figure} [t]
\begin{center}
\begin{tabular}{c}
\includegraphics[height=11.0 cm]{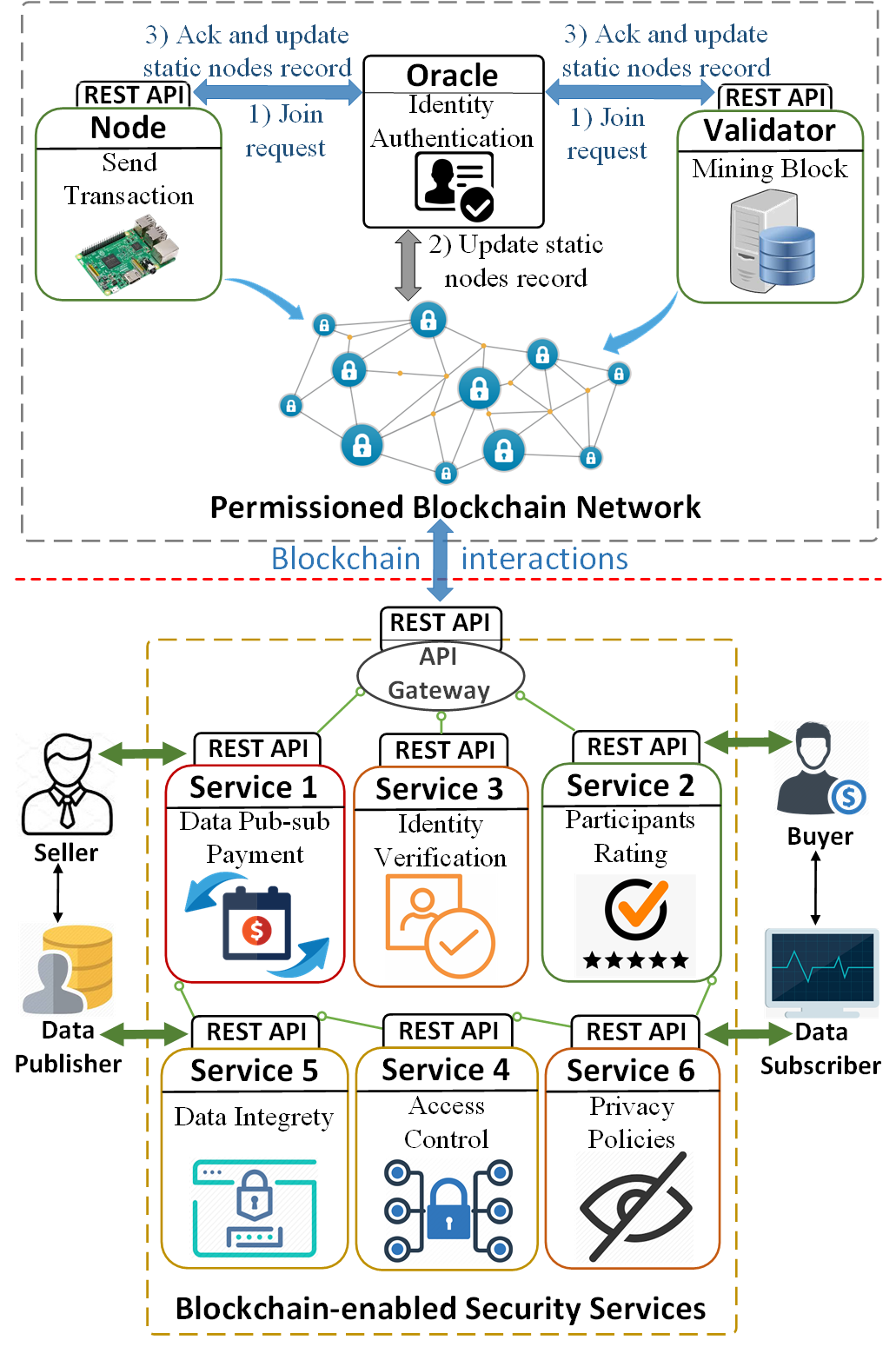}
\end{tabular}
\end{center}
\caption[example] { \label{fig:I3_architecture} Illustration of BlendSM-DDM System Architecture.}
\vspace{-10pt}
\end{figure}

Inspired by blockchain and containerized microservices technology, BlendSM-DDM is proposed as a decentralized security services architecture for DDM system, which is shown in Fig. \ref{fig:I3_architecture}. Trinity \cite{ramachandran2019trinity} distributes the pub-sub broker using the blockchain technology to prevent byzantine faults. A similar approach could be adopted to distribute the microservices among consensus nodes. A permissioned blockchain infrastructure running on a decentralized peer-to-peer network is managed by DDM administrators, and it could provide a reliable, traceable and immutable distributed ledger for trusted data sharing services in DDM system. Utilizing the containerized microservices architecture, the key elements and security features of DDM system are decoupled into multiple microservices and deployed on distributed computing hosts. Each microservice exposes REST APIs to accept service requests, and API gateway could be managed by either DDM administrator or service providers. Those decentralized microservices work as a blockchain-enabled services pool to offer a scalable, flexible and lightweight security and privacy protection mechanism for data marketplace. 

\subsection{Permissioned Blockchain Network}
The permissioned blockchain network in the BlendSM-DDM requires that only registered entities are authorized access to the network, so that they can interact with the blockchain and perform blockchain-based services, such as mining block, sending transactions and deploying smart contracts. The upper part of Fig. \ref{fig:I3_architecture} illustrates the identity authentication process to enroll a new participant in the permissioned blockchain network. Given computation capacity of host devices and management policy, all participants are categorized as validator (miner) or node (non-miner). The validators are able to cooperatively executing consensus algorithm to maintain the integrity and immutability of distributed ledger. While nodes are only capable of launching transactions to synchronize data with the blockchain. 

To join the permissioned blockchain network, the participants must send joining requests to an oracle for registration and identity authentication. The administrator of DDM system could work as such oracle, and maintain a global identity profile and authorization policies for the permissioned blockchain network management. After approved by the oracle, the new entity's node information will be added to a global static node record, and will be broadcasted to all certificated participants in the permissioned blockchain network accordingly. The membership revocation occurs when an entity explicitly launches a leaving request or the oracle implicitly rule out any misbehaved nodes.

Compared to a public blockchain network that suffers from performance and security issues, such as low throughput, poor privacy and lack of governance, the permissioned blockchain can enforce a more efficient but lightweight consensus mechanism to improve the transaction rate and data storage capacity. Furthermore, through limiting participants with clearly defined privacy policies, the sensitive data could be only visible to the authorized participants but not to the public. Therefore, the permissioned blockchain network also enable a more strict privacy policy than a public blockchain network does.

\subsection{Blockchain-enabled Security Services}

The blockchain-enabled security services layer, as shown in lower part of Fig. \ref{fig:I3_architecture}, acts as a fundamental microservices oriented infrastructure to support business logic functions of the DDM system as well as security mechanism. The key elements and operations are described below.

    \noindent \emph{1. Data Pub-sub Payment}: The seller and buyer use their blockchain addresses as unique IDs to join the data exchange and payment activities. The cryptocurrency based micro-payments protocol is designed to collect data in exchange for payment or other incentives. The key transaction records, such as invoices and receipts, are stored in a distributed ledger, so that all participants could verify the transactions. 
    
    \noindent \emph{2. Identity Verification}: Since each blockchain account is uniquely indexed by its address that is derived from his/her own public key, the DDM administrator could deploy smart contract enabled microservices, which expose RESTful API to other microservices-enabled providers for referring identity verification results. 
    
    \noindent \emph{3. Participants Rating}: To evaluate the quality of the data product provided by a given seller and reliability of payment by a buyer, rating smart contract based microservices are developed and deployed by trusted agents of the DDM system to allow buyers and sellers to rate each other. Owing to data transparency and immutability in the distributed ledger, the rating records stored in the blockchain network cannot be falsified or tampered with. The rating information would be used as recommended references for future potential buyers and sellers. 
    
    \noindent \emph{4. Data Integrity}: Data integrity technologies are mainly to ensure data access at the same time avoid storing a huge amount of data in the blockchain. The data integrity microservices provides the dynamic data synchronization and efficient verification through a hashed index authentication process. The data publishers just simply interact with authorized ABI functions of the smart contract to save the hashed index of data to blockchain. In the hashed index verification process, data subscribers just query a hashed key-value index from the blockchain and compares it with calculated hash values of the received data.   
    
    \noindent \emph{5. Access Control}: The data owners could transcode access control models and policies into a smart contract-based access control (AC) microservice. The AC microservices allows those service providers to control their data and resource without relying on a centralized third party authority, and it provides a decentralized access control solution to the DDM system.
    
    \noindent \emph{6. Privacy Policies}: The privacy microservices is mainly for privacy-sensitive data management, so that the privacy-sensitive data is not linked or accessed by unauthorized entities. Secure communications using encryption and saving sensitive data as a hash value mapping to actual data are efficient solutions to protect privacy. Furthermore, properly designing access control policies could effectively prevent unauthorized access to sensitive data.
   


\section{Conclusions}
\label{sec:conclusion}  
In this paper, BlendSM-DDM, a blockchain-enabled decentralized security microservices framework is proposed to ensure data exchanges and payments in DDM systems like the Intelligent IoT Integrator (I3). The comprehensive overview of the system architecture is introduced and key elements are illustrated. The BlendSM-DDM could offer a scalable, flexible and lightweight security and privacy protection mechanism for business operations in DDM system. In ongoing work, we are designing the various elements presented in the architecture, and implementing a prototype running on current DDM system to verify the feasibility of the proposed solutions.

\bibliographystyle{IEEEtranS} 
\bibliography{report} 

\end{document}